\documentclass[twocolumn,PRB,amsmath,amssymb]{revtex4-1}
\usepackage{amssymb}
\usepackage{amsmath}
\usepackage{graphicx}
\usepackage{epstopdf}
\usepackage{bm}%
\usepackage{gensymb}
\usepackage{soul}
\usepackage{sidecap}
\usepackage[normalem]{ulem}
\usepackage{epsfig}
\usepackage{graphicx}
\usepackage{amsfonts,amssymb}
\usepackage{amsmath}
\usepackage{color}
\usepackage {times}
\usepackage{epstopdf}

\newcommand{\bl}{\begin{aligned}}
\newcommand{\el}{\end{aligned}}

\newcommand{\be}{\begin{equation}}
\newcommand{\ee}{\end{equation}}   
\newcommand{\bea}{\begin{eqnarray}}
\newcommand{\eea}{\end{eqnarray}}
\newcommand{\ba}{\begin{array}}
\newcommand{\ea}{\end{array}}

\newcommand{\q}{{\bf q}}
\renewcommand{\k}{{\bf k}}
\newcommand{\Q}{{\bf Q}}

\graphicspath{{Figs/}}

\begin{document}

\title{Thermal evolution of single-particle spectral function in the half-filled Hubbard model and pseudogap}

\date{\today}
\author{Harun Al Rashid}
\author{Dheeraj Kumar Singh}
\email{dheeraj.kumar@thapar.edu }
\affiliation{School of Physics and Materials Science, Thapar Institute of Engineering and Technology, Patiala-147004, Punjab, India}

\begin{abstract}
In the half-filled one-orbital Hubbard model on a square lattice, we find pseudogap-like features in the form of two-peak structures associated with the momentum-resolved spectral function. These features exist within the temperature window $T_N \lesssim T \lesssim T^*$, where $T_N$ is the N\'{e}el temperature and $T^*$ is the temperature below which there exists a well-formed dip in the density of state. Inside the window $T_N \lesssim T \lesssim T^*$, the peak-to-peak separation in the two-peak structure of the momentum-resolved spectral function grows on moving away from the point ($\pi/2, \pi/2$) along the normal state Fermi surface towards $(\pi, 0)$, a behavior with remarkable similarities to what is observed in the  $d$-wave state and pseudogap phase of high-$T_c$ cuprates. We unveil these features by using a parallelized cluster-based Monte-Carlo method for simulating the magnetic order parameter fields on a superlattice. The method enables us to access the momentum-resolved single-particle spectral function corresponding to a lattice size of $\sim$ 240 $\times$ 240 with almost negligible finite-size effect. 

\end{abstract}

\pacs{}
\maketitle
\newpage
\section{Introduction}

The mechanism of charge-carrier localization in  complex correlated-electron systems, associated with the  phase transition from the metallic to insulating state, has been a recurrent theme despite a significant theoretical and experimental progress made in the last few decades~\cite{imada,lee}. This is largely because several aspects of the metal-insulator transition (MIT), including the evolution of the momentum-dependent spectral features of the quasiparticle excitations with temperature, are not very well understood~\cite{kim}. Mostly observed in the correlated $d$ and $f$ electron systems~\cite{hotta}, MIT may depend on a variety of factors enhancing the complexity of the nature of transition. The factors are strong electron-electron interaction~\cite{mott,george,kotliar}, lowering of translational symmetry due to a phase transition as in the case of paramagnetic (PM)-to-antiferromagnetic (AFM) transition\cite{slater}, trapping of charge carriers because of disorder caused by the impurities~\cite{anderson,evers}, polaron formation in the presence of strong electron-phonon coupling etc~\cite{kim1,kim2}.  

MIT in a correlated-electron system does not necessarily require a transition to a phase with reduced symmetry such as the one with an antiferromagnetic arrangement of spins~\cite{george}. In these systems with only one orbital, which is half filled, the energy cost of double occupancy is large. Therefore, the delocalization of electrons becomes increasingly suppressed upon lowering of the temperature. The corresponding signature is noticed as a dip in the density of states (DOS) near the low-energy region. The dip is a consequence of the spectral weight transfer from the low- to high-energy region as the electrons get increasingly localized. Subsequently, two incoherent broad peaks separated by an energy scale comparable to the electron-electron interaction energy are formed as originally proposed by Mott \textit{et al}~\cite{mott}.

Within the Landau fermi-liquid theory, a quasiparticle peak centered at the Fermi level is expected to exist above the MIT temperature in the systems with a weak-to-moderate correlations~\cite{brinkman}. The quasiparticle peak, often referred to as Brinkman-Rice (BR) peak, is expected to get increasingly narrower on approaching MIT and disappear finally. Recent experiments does provide the signature for the existence of three-peak structure of the quasiparticle spectra in several transition-metal oxides including V$_2$O$_3$~\cite{mo,panaccione,eguchi,rodolakis}, SrVO$_3$~\cite{sekiyama}, CeTO$_{3+\delta}$ etc~\cite{akaki}. On the other hand, in the strongly correlated systems such as high-$T_c$ cuprates~\cite{damascelli}, titanates~\cite{fujimori} etc., the low-energy coherent quasiparticle peak may be absent.   

Considerable details of the quasiparticle spectra have been obtained using spectroscopies based on the optical excitations~\cite{uchida, marel,qazilbash} and photoemission~\cite{eguchi,rodolakis,matsuura,steve, dhaka}. However, there is not enough progress in understanding the momentum-resolved spectra across the Mott transition using angle-resolved photoemission spectroscopy (ARPES). There exists only a few studies mostly in the parent compound of high-$T_c$ cuprates such as Ca$_2$CuO$_2$Cl~\cite{ronning} and Sr$_2$CuO$_2$Cl~\cite{kim3}. ARPES measurements in these compounds both below and above $T_N $ indicate a $d$-wave like form $|\cos k_x - \cos k_y|$ of the gapstructure along the Fermi surface of strange metallic state. These signatures were suggested to be  indicative of a possible link to the $d$-wave superconductivity.

The high-$T_c$ cuprates does not exhibit fermi-liquid behavior at higher temperature for a small number of holes irrespective of the low-temperature phase being either $d$-wave superconductivity or AFM state~\cite{damascelli}.
 The widely accepted temperature-vs-doping phase diagram of cuprates contains nearly a straight line ~\cite{vishik} boundary separating the pseudogap (PG) phase from a metallic phase with non-fermi liquid behavior. It may be noted that the straight line is located above the curve separating the PG phase and N\'{e}el order. This indicates that the PG phase is also precursor to the N\'{e}el order. In that case, how does the single-particle spectral function evolve as a function of temperature above N\'{e}el order? How can an intermediate phase be precursor to two distinct phases which may be competitor/hostile to each other? Does the spectral gap above $T_N$ resemble that of the PG or is it similar momentum-dependent gap structures seen in the undoped cuprates. The answers to these questions may also throw light on the origin of PG phase and associated features. 
 
We search answers to the above questions, by examining temperature dependent single-particle spectral function within one-band Hubbard model, which has been explored earlier with variety of methods~\cite{huscroft,stanescu,senechal,wu}. While the dip in the DOS was shown to exist, the behavior of momentum-dependent gap-structure above $T_N$ remains  largely unknown because of difficulties involved in incorporating spatial correlations in some of the methods. In this paper, we use exact-diagonalization + Monte-Carlo (ED+MC) based scheme which is one of the most suitable approaches to study spatial and thermal fluctuations. However, this method like others is often accompanied with inaccessibility to a larger size system~\cite{mayr,anamitra}, which is unfavorable for a good momentum resolution. The difficulty increases further due to a significant broadening of the  quasiparticle peaks  when the low-temperature phase is N\'{e}el order. 

We use a methodology which combines three techniques to overcome the size-dependent limitation thus enabling one to access the momentum-resolved spectral function almost free of any finite-size effect. The momentum-resolved spectral function with a high resolution is prerequisite in order to conclusively establish the nature of the spectral function particularly at higher temperatures where the broadening caused by the spatial and thermal fluctuations dominates. Three techniques are traveling-cluster approximation (TCA), parallelization of update process and twisted-boundary conditions (TBC) to be discussed later.  

We arrive at the following results obtained by adopting the above mentioned approach: (i) The spectral function $A(\k, \omega)$ shows  significant broadening owing to large fluctuations in the order parameter fields at higher temperatures. (ii) As one approaches $T_N$ from below, the dip in the DOS takes the $V$-shape like one observed in the case of $d$wave superconductivity. The $V$-shaped  dip continues to be present in the vicinity of $T_N$ and beyond through it becomes increasingly shallow. 
(iii) There exists a temperature scale $T^*$ beyond which, the dip in the DOS becomes very small, doesn't show sign of any further change on increasing temperature, and it persists even beyond. 
(iv) Within the temperature window $T_N \lesssim T \lesssim T^*$, the spectral function $A(\k, \omega)$ has a two peak structure with a dip at $\omega = 0$ for all the momenta along the normal state Fermi surface. (v) The peak-to-peak separation, which is nearly constant below $T_N$, rises on moving from ($\pi/2, \pi/2$) to ($\pi$,0), a behavior similar to $d$wave superconductivity and PG phase. (vi) The nearest-neighbor and next-nearest neighbor magnetic correlations are not negligibly small beyond $T_N$.

\section{model and method}
\subsection{Model}
We consider Hubbard model on a square lattice given by
\begin{equation}
{{\mathcal H}_{}}=\sum_{\langle ij\rangle\sigma}(t_{ij} - \delta_{ij} \mu) 
d^{\dagger}_{i\sigma}d^{}_{j\sigma}  + U\sum_{i}n_{i\uparrow}n_{i\downarrow},
\end{equation}
where $d^{\dagger}_{i\sigma}$ ($d_{i\sigma}$) is the creation (destruction) operator for an electron at site $ i$ with spin $\sigma$. $t_{ij} = t$ is the hopping parameter between two neighboring sites $i$ and $j$. The second term denotes the on site Coulomb repulsion between two electrons having opposite spins. The energy scale in this work is set in the unit of $t$ throughout. The chemical potential $\mu$ is chosen in order to keep the band filling fixed at half filling.  

\subsection{Methodology}
In the simulation process, the effective Hamiltonian can be obtained via a Hubbard-Stratonovich transformation, which is equivalent to meanfield-like decoupling of the Hubbard term in the magnetic channel, if the magnetic vector fields ${\bf m}_is$, to be defined below, are treated as classical fields. This approximation ignores the temporal fluctuations of the vector fields but allows for the spatial fluctuations, yields
\begin{equation}
{H}_{m} 
= - \frac{U}{2} \sum_{i \sigma} \Psi^{\dagger}_{i}({\bf m}_{i}\cdot\vec{\sigma}) \Psi_{i}
+ \frac{U}{4}\sum_{i}{\bf m}_{i}^{2}
\vspace{-3mm}
\end{equation}
with $\Psi^{\dagger} = (d^{\dagger}_{i \uparrow}, d^{\dagger}_{i \downarrow} )$. The $j$-th component of magnetic order-parameter field at site $i$ is ${ m}^{j}_i = \frac{1}{2} \langle \Psi^{\dagger}_{i}{\sigma}^{j} \Psi_{i} \rangle$, where ${\sigma}^{j}$ is $j$-th component of Pauli matrices. Note that the electronic part of the Hamiltonian 
\begin{equation}
 H_{el} =  \sum_{\langle ij\rangle\sigma}(t_{ij} - \delta_{ij} \mu) 
d^{\dagger}_{i\sigma}d^{}_{j\sigma}- \frac{U}{2} \sum_{i \sigma} \Psi^{\dagger}_{i}({\bf m}_{i}\cdot\vec{\sigma}) \Psi_{i} 
\end{equation}
can be readily diagonalized provided that the field configurations $\{{\bf m}_i\}$ are given. Thermally equilibrated $\{{\bf m}_i\}$s, in turn, are generated with the help of MC sampling according to the distribution 
\begin{eqnarray}
P\{{\bf m_i} \} &\propto& {\rm Tr}_{d,d^{\dagger}} e^{-\beta H_{eff}},
\end{eqnarray}
where  the effective Hamiltonian $H_{eff} = H_{el}\{{\bf m}_i\} + H_{cl}$ with $H_{cl}= (U/4)\sum_i {\bf m}_i^2$. The accuracy of this MC-based approach increases with a rise in temperature as the thermal fluctuations become increasingly dominant. The approach reduces to the standard Hartree-Fock approximation in the low-temperature limit.

First, we use a cluster-based approximation~\cite{kumar} for the MC update, which instead of Hamiltonian corresponding to full lattice $N_L \times N_L$ , considers the Hamiltonian of a smaller lattice size $N_c \times N_c$ centered at the update site for each update. This is a reasonable assumption given the fact that the effect of the update shall fall quickly on moving away from the update site~\cite{kohn} and it  reduces the computational cost significantly. 

Second, an additional reduction of computational cost in the simulation process is achieved through parallelization of the update process. In the standard simulation process, an update site is chosen, a change is proposed in the classical field, energy of the system is calculated, and the update is accepted/rejected according to the Metropolis algorithm. In the parallelized update scheme, more than one update sites are chosen at the same time when $N_p$ number of processors are available.  $N_L \times N_L$ lattice is divided into blocks of size $N_b \times N_b$ so that $N_L$ is divisible by $N_b$. $N_p$ identically positioned update sites are chosen from the successive blocks and corresponding cluster Hamiltonians are diagonalized, energies are calculated, and the fields are updated according to Metropolis algorithm. For the next update, the process is repeated for next $N_p$ successively placed set of blocks and so on.

Third, we use twisted-boundary conditions (TBC)~\cite{salafranca} which corresponds to a superlattice formed by repeating thermally equilibrated fields configuration for a lattice size $N_L \times N_L$ obtained earlier $-$ $N_t$ number of times in both $x$ and $y$ directions. Then, one can study the spectral features of a lattice size $N_LN_t \times N_LN_t$. Throughout in the current work, $N_c = 8$, $N_L = 40$ and $N_t = 6$ are used unless otherwise stated, which corresponds to an effective lattice size  $240 \times 240$.

\subsection{Onset temperature for magnetic order}
The signature of onset of long-range magnetic order is obtained from the thermally averaged structure factor given by
 \be
 S({\bf Q}) = \frac{1}{N^2}\sum_{i,j} \langle {\bf m}_i \cdot {\bf m}_j \rangle e^{i {\bf Q}\cdot({\bf r}_i - {\bf r}_j)}.
 \ee
 Here, ${\bf Q} = (\pi, \pi)$ for the checkerboard-type antiferromagnetic order and ${\bf r}_i $ is the position of vector field ${\bf m}_i$. The temperature, at which the onset of rise in the structure factor $S({\bf Q})$ during the annealing process occurs, is taken as the transition temperature $T_N$.
 
 It has been widely speculated that the pseudogap features may be a consequence of short-range magnetic correlations. The onset of short-range magnetic correlations can be studied with the help of functions given by
 \be
 \phi_i = \frac{1}{4N}\sum_{\langle {i,j} \rangle } |\langle {\bf m}_i \cdot {\bf m}_j \rangle|,
 \ee
 where ${\langle {i,j} \rangle }$ denotes summation over nearest neighbors for $\phi_1$, next-nearest neighbors for $\phi_2$ etc. Note that the factor of four appears in the denominator as there are only four nearest or next-nearest neighbors for a square lattice.  
\begin{figure}[t]
\begin{center}
\vspace{2mm}
\includegraphics[scale = 1.0,angle = 0]{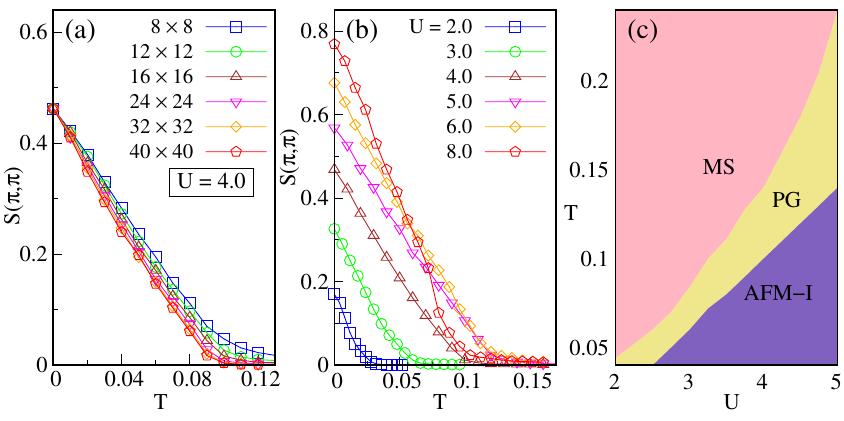}
\vspace{-5mm}
\end{center}
\caption{(a) The AFM structure factor $S(\Q)$ as a function of temperature for different lattice sizes $N_{L} \times N_{L}$ with $N_L =$ 8, 12, 16, 24, 32 and 40. (b) $S(\Q)$ for different $U$s indicating the onset of long-range AFM correlation. (c) $T-U$ phase diagram based on the transition temperatures ($T_N$ and $T^*$), where metallic (MS), pseudogap (PG) and AFM-insulating phases are shown.}
\label{f1}
\end{figure}
\begin{figure}[t]
\begin{center}
\hspace{-0mm}
\includegraphics[scale = 1.0,angle = 0]{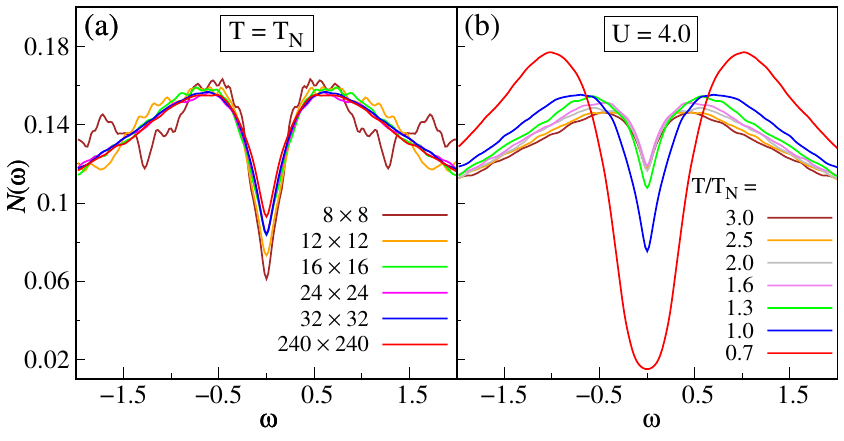}
\vspace{-10mm}
\end{center}
\caption{(a) DOS at $T_N$ for different lattice sizes $N_{L} \times N_{L}$ with $N_{L} =$ 8, 12, 16, 24, 32 and 40. For, $N_L = 40$,  twisted-boundary condition is used to obtain DOS for an effective lattice size of 240 $\times$ 240 (b) DOS at different temperatures for $U = 4.0$. The dip in the DOS, indicative of PG, can be seen to persist even beyond $T_N$.}
\label{f2}
\end{figure}

\begin{figure}[t]
\begin{center}
\hspace{-0mm}
\includegraphics[scale = 0.27,angle = 0]{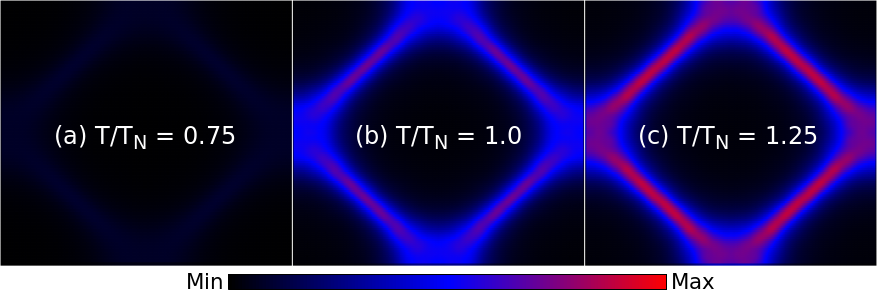}
\vspace{-7mm}
\end{center}
\caption{Spectral function ${A}({\bf k}, \omega)$ for $\omega = 0$ and different temperatures (a) $T/T_N = 0.75$, (b) $1.0$ and (c) $1.25$. The range of ${A}({\bf k}, \omega)$ is common to all three figures while the range of $k_x$ and $k_y$ is [$-\pi, \pi$]. There is significant broadening   throughout along the Fermi surface and at ($\pi, 0$) point in particular. ${A}({\bf k}, 0)$ appears to have comparatively larger values near ($\pi/2, \pi/2$) in comparison to ($\pi, 0$).}
\label{f3}
\end{figure}
\subsection{Gap-structure determination}
 In order study DOS below and beyond $T_N$, we use TBC, once the thermal equilibration is reached. For instance, $t_{{\bf i} {\bf j}} \rightarrow t_{{\bf i} {\bf j}} e^{- i (q_x a_x + q_y a_y)}$, where $q_x,q_y = 0, 2\pi/N_t, 4\pi/N_t,...2\pi(N_t-1)/N_t$ and $N_t$ is the number of lattices in the superlattice \textit{i.e}, number of repetition along $x$ and $y$ directions of the lattice  under consideration in the parallelized MC scheme. In our calculation, $N_t  = 6 $ while setting the  lattice constants $a_x = a_y = 1$.
Then, the DOS is calculated as  
\be
N(\omega)  = \sum_{\q, \lambda, {\bf i}} |{\psi}_{\q,\lambda}({\bf i})|^2\delta
(\omega - E_{\q , \lambda}) 
\ee
where $E_{\q, \lambda}$ are the eigenvalues of the fermionic part of the full Hamiltonian obtained. $|{\psi}_{\q,\lambda}\rangle$ is the eigenvector of the Hamiltonian. 

By using the twisted-boundary condition, the single particle spectrum is calculated as 
\be
A(\k, \omega)  =  \sum_{\q, \lambda} ( |\langle \k  
|{\psi}_{\q,\lambda}\rangle |^2 \delta(\omega - E_{\q, \lambda}) , 
\ee
where
\be
\langle \k|{\psi}_{\q,\alpha}\rangle = \sum_l \sum_i \langle \k |
l, i \rangle \langle l, i | {\psi}_{\q,\lambda} \rangle.
\ee
$l$ is the sublattice index and $i$ is a site index within the sublattice.
\section{Results}
 For our calculation, we have chosen $U = 4$, which yields the exchange coupling $J \approx 4t^2/U = 1$ in the unit of $t$, in accordance with recent estimates~\cite{jang}. However, as discussed later, we find that the pseudogap features get enhanced with increasing $U$. We start MC simulations at a temperature $T \sim 0.2$ or even beyond, which is more than twice the magnetic transition temperature $T_N$, and reduce the temperature to cool down the system in the steps of $T \sim 0.01t$. A small temperature step is considered in order to bypass any metastable states during entire cooling process. 
 
  To see that the $S({\bf Q})$ for $N_L = 40$ is nearly free from finite-size effect, it is plotted as a function of temperature for different lattice sizes as shown in figure~\ref{f1}(a). $S({\bf Q})$s approach the same value asymptotically in the limit $T\rightarrow 0$, however, a size dependence can be noticed near $T_N$. In particular, $T_N$ shows a small reduction with an increase in the lattice size, while for $N_L = 40$ in the current work, it appears to nearly coincide with its asymptotic value in the large $N_L$ limit. 
 
 Fig.~\ref{f1}(b) and (c) show the dependence of $T_N$ on the Coulomb repulsion parameter $U$. First, the magnetic moment size increases with $U$ and approaches towards its saturation value. Second, $T_N$ rises almost linearly with an increase in $U$ up to $U \sim 6$, where it maximizes, and then decreases beyond. Fig.~\ref{f1}(c) shows an additional temperature scale $T^*$, which corresponds to a temperature where the dip in the DOS becomes shallow enough and there is no further change upon increase in temperature. The shallow dip continues to exist even beyond $T^*$ up to a very large temperature as revealed in our calculation, which may perhaps be the consequence of persistent short-range correlations. Unlike $T_N$, $T^*$ exhibits a sharp rise  as $U$ increases.
 
One of the earliest experimental signatures of the PG was a dip obtained in the DOS, which was found to persist above the superconducting-transition temperature obtained~\cite{renner1}. Such a dip in the DOS above superconducting transition is found to exist for a wide range of hole doping where the PG phase is observed.

 As pointed out above, $T^*$ correspond to the temperature when the dip in the DOS has become shallow and there is no further change with an increase in the temperature. This occurs near $T \sim 1.5T_N$ for $U = 4$ (Fig.~\ref{f2}(b)). In order to show that the result is free of finite-size effect, the size dependence of DOS at $T=T_N$ is also examined for $U = 4$, where the absence of oscillation in the DOS originating from finite-size effect and shallowest dip can be noticed for a larger lattice (Fig.~\ref{f2}(a)). The dip in the DOS takes the $V$-shape for $T \sim T_N$, which continues to exist beyond $T_N$ in a way similar to the gap persisting beyond superconducting transition in the underdoped cuprates. Next, we address the question whether $V$-shaped dip in the DOS beyond $T_N$ has  momentum-dependent gap structure similar to the PG phase with the help of spectral function $A( {\bf k}, \omega)$. 
 \begin{figure}[t]
\begin{center}
\hspace{-0mm}
\includegraphics[scale = 1.0,angle = 0]{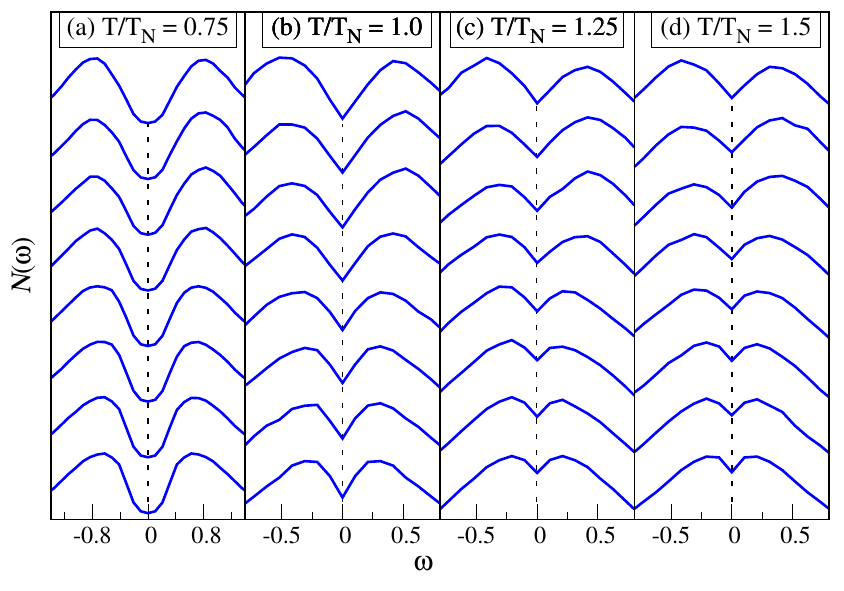}
\vspace{-12mm}
\end{center}
\caption{$\mathcal{A}({\bf k}, \omega)$ as a function of $\omega$ at different temperatures $T/T_N = $ (a) 0.75 (b) 1.00, (c) 1.25 and (d) 1.50 along the Fermi surface. The curves at the bottom and top corresponds to $(\pi/2, \pi/2)$ and $(\pi, 0)$, respectively, while the others to the points in between as one moves from $(\pi/2, \pi/2)$ to $(\pi, 0)$. At $T/T_N$ = 0.75, there is no change in the gap profile and size as one moves from $(\pi/2, \pi/2)$ towards $(\pi, 0 )$, while the same is not true for $T/T_N \gtrsim 1$. }
\label{f4}
\end{figure}
\begin{figure}[]
\begin{center}
\hspace{0mm}
\includegraphics[scale = 1.0,angle = 0]{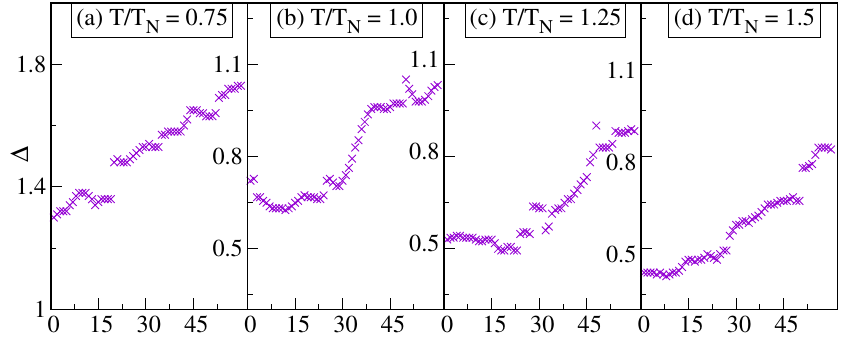}
\vspace{-8mm}
\end{center}
\caption{Peak-to-peak separation in the spectral function ${A}({\bf k}, \omega)$ as one moves along the Fermi surface away ($\pi/2, \pi/2$) to ($\pi, 0$) at different temperatures corresponding to $T/T_N = $ (a) 0.75, (b) 1.00, (c) 1.25 and (d) 1.50. 0 and 60 corresponds to $(\pi/2, \pi/2)$ and $(\pi, 0)$, respectively. The peak-to-peak separation $\Delta$ rises on moving from ($\pi/2, \pi/2$) towards ($\pi$, 0) for $T \gtrsim T_N$. The fluctuations in $\Delta$ is a consequence of significant broadening as temperature is increased.}
\label{f5}
\end{figure}

Fig.~\ref{f3} shows the $A({\bf k}, 0)$ at different temperatures while keeping its range same. For $T < T_N$, $A({\bf k},0)$ is vanishingly small all along the normal state Fermi surface. It increases on approaching $T_N$ and becomes non-negligible at $T_N$, which doesn't indicate, however, its gaplessness. For temperatures near $T_N$ and beyond, a significant broadening can be seen when compared to that of the Bogoliubov quasiparticles arising due to thermal phase fluctuations beyond the superconducting transition temperature. The broadening obtained here is devoid of any artifact which may result from the finite-size effect. This directly follows from the momentum resolution in our calculation, which is of the order $\Delta k_x = \Delta k_y \sim 2.5 \times 10^{-2}$, in the unit of $1/a$, significantly small to cause any such artifact. 
\begin{figure}[]
\begin{center}
\hspace{0mm}
\includegraphics[scale = 1.0,angle = 0]{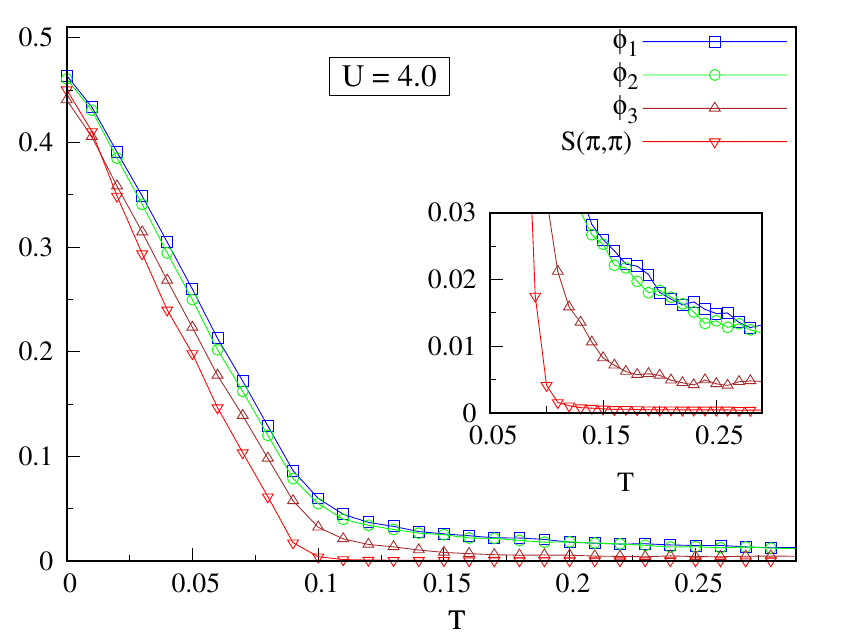}
\vspace{-10mm}
\end{center}
\caption{$\phi_1$, $\phi_2$, and $\phi_3$ denote  nearest, next-nearest, and next-next-nearest neighbor magnetic correlations, respectively. For comparison, the long-range magnetic correlation function $S(\Q)$ is also included.}
\label{f6}
\end{figure}
There is a relatively enhanced thermal broadening towards ($\pi, 0$) because of the availability of a large phase space. In addition, $A(\k, 0)$ increases as one move towards $(\pi/2, \pi/2)$ along the normal state Fermi surface, a feature also observed in the pseudogap phase. However, additional definitive information about the nature of gap structure above $T_N$ can only obtained by examining $A({\bf k}, \omega)$ for different momenta along the Fermi surface. 

Fig.~\ref{f4} shows the momentum-resolved spectra as a function of energy $\omega$ for temperatures below and above $T_N$. Top and bottom curves in each subfigure are $A({\bf k}, \omega)$ for the Fermi points ($\pi/2, \pi/2$) and ($\pi, 0$), respectively, while other curves are for the points in between. A well-developed gap in $A({\k}, \omega)$ at $T=0.75T_N$ is on expected line, which appears to be robust in size when moving along the normal state Fermi surface. However, two distinct features may be noticed in the vicinity of $T_N$ and beyond. First, the gaps in $A({\k}, \omega)$ is increasingly being filled up on moving away from ($\pi,0$) to ($\pi/2, \pi/2$). Second, the gap size, \textit{i.e.}, peak-to-peak separation decreases at the same time. These features build up with increasing temperature. 

Next, we look at the dependence of peak-to-peak separation $\Delta$ as a function of distance from ($\pi/2, \pi/2$) along the normal state Fermi surface (Fig.~\ref{f5}). We find that there is an overall increase in $\Delta$ irrespective of temperature beyond or below $T \sim T_N$. In particular, at $T = T_N$, $\Delta_p$ is non-vanishing at ($\pi/2, \pi/2$) and it increases on moving towards ($\pi, 0$) and gets nearly doubled, although an overall reduction in its value is noticed as temperature increases. The fluctuations in $\Delta$ when one moves along normal state Fermi surface is a consequence of a large  broadening resulting from thermal and spatial fluctuations in the orientation of magnetic moments. 

The spectral features found above appear to arise from the short-range magnetic correlations which persist beyond $T_N$. Fig.~\ref{f6} shows the plots of the nearest ($\phi_1$), next-nearest ($\phi_2$), and next-next-nearest neighbor ($\phi_3$) magnetic correlation functions together with the long-range magnetic correlation function $S(\Q)$ as a function of temperature.  As it can be noticed, $S(\Q)$ vanishes completely and $\phi_3$ is also very small beyond $T_N$, whereas both $\phi_1$ and $\phi_2$ don't vanish and show nearly constant value $\sim$ 0.01, which is of an order similar in magnitude to that of $S(\Q)$ near the onset of long-range correlation. Presence of short-range magnetic correlations can be noticed up to the highest temperature $T\sim 3T_N$ considered in this work.
\section{discussion}
The nature of single-particle spectral function, obtained here across the MIT in the Hubbard model at half filling, is able to capture the essential spectral features observed in the ARPES measurement of parent compound of high-$T_c$ cuprates~\cite{ronning,kim3}. On the other hand, the spectra exhibit similarities as well as dissimilarities with what have been observed in the $d$-wave state and PG phase~\cite{renner1,ding,loeser,norman,yoshida,kanigel1,kanigel2,hasimoto,dheeraj}. Increase in the peak-to-peak separation as one moves towards $(\pi, 0)$ from $(\pi/2, \pi/2)$ along the Fermi surface is a feature similar to the $d$-wave state and PG phase. However, the separation vanishes at the point ($\pi/2, \pi/2$) in the case of $d$-wave state and along an arc centered at ($\pi/2, \pi/2$) in the PG phase. In contrast, the peak-to-peak separation is non zero in the pseudogap-like state found beyond $T_N$ in the current work. Furthermore, the dip in the DOS is shallow and therefore it is similar to what is observed in the PG phase unlike $d$-wave superconductivity with fully formed $V$-shaped gap.

 It is widely believed that the persistent short-range magnetic correlations may be the key factor responsible for the momemntum-dependent spectral features observed in the underdoped cuprates beyond $T_N$. Our findings do lay support to this idea. Experimentally, the onset temperature of both N\'{e}el order and the pseudogap phase increases on approaching zero doping, which indicates a strong link between the pseudogap-like spectral features and the  magnetic correlation, as the latter maximizes in the parent compound~\cite{vishik}. Therefore, the pseudogap-like features of the gapstructure existing beyond $T_N$ are expected to be qualitatively similar in the cases with and without the long-range hoppings. The difference that may arise is the asymmetry in the spectral features introduced by the broken particle-hole symmetry when the long-range hopping is considered. Secondly, the $T_N$ will be relatively smaller for a given value of $U$ because of the frustration brought in by the long-range hoppings.

In this work, we were primarily focused on  establishing the nature of momentum-resolved single-particle spectral function across the MIT. For that purpose, we chose the repulsive Coulomb interaction $U = 4$, a value less than the unscreened one, which agrees with the recent works~\cite{jang}. Since $U < W$, where $W$ is the bandwidth, the BR peak is expected to exist beyond the MIT temperature. 
Within our treatment of the meanfield, temporal fluctuations are ignored as the fields are treated classically in order to obtain the spectral functions with a high-momentum resolution at the cost of the peak. On the other hand, our calculation incorporates both the thermal and spatial fluctuations in the fields in order to conclusively establish the nature of spectral features existing beyond $T_N$. 
The BR peak together with incoherent spectra at high energy is captured by the dynamical mean-field theory (DMFT)~\cite{george,kotliar}, wherein the dynamical nature of meanfield is incorporated. This is, however, at expense  of spatial fluctuations, as the method ignores the momentum dependence of self-energy, and the local part of the self-energy is treated as an impurity problem with a dynamical mean-field. 
The BR peak, though small near the MIT temperature, may also affect the momentum-resolved spectra. Nonetheless, the results obtained here provide important insights into the thermal evolution of momentum-resolved spectral function in the strongly-correlated systems including those of the high-$T_c$ cuprates which exhibit non-fermi liquid behavior. 
\section{conclusions}
In conclusions, we have mapped out the momentum-dependent gap structure of single-particle spectral functions in the half-filled  Hubbard model across the N\'{e}el temperature by using a method which provided us access to a very large lattice size with almost negligible finite-size effect. We obtained spectral features which show several remarkable similarities to those observed in the $d$-wave or pseudogap phase of high-$T_c$ cuprates thus indicating that the pseudogap may originate from the short-range magnetic correlations. Although shown for a single-orbital Hubbard model, these features are expected to be observed in other correlated systems in the absence of additional competing ordering tendencies, which may otherwise suppress the persistent short-range magnetic correlations responsible for these features.

\section*{Acknowledgement}
D.K.S. was supported through DST/NSM/R\&D\_HPC\_Applications/2021/14 funded by DST-NSM and start-up research grant SRG/2020/002144 funded by DST-SERB. He would also like to thank Y. Bang and Pinaki Majumdar for useful comments and suggestions.


\begin{thebibliography}{100} 



\bibitem{imada} M. Imada, A. Fujimori, and Y. Tokura, Rev. Mod. Phys. {\bf 70}, 1039 (1998).

\bibitem{lee} P. A. Lee, N. Nagaosa, and X.-G. Wen, 
Rev. Mod. Phys. {\bf 78} 17 (2006).


\bibitem{kim} S. Y. Kim, M.-C. Lee, G. Han, M. Kratochvilova, S. Yun, S. J. Moon, C. H.  Sohn, J.-G. Park, C. Kim, and T. W. Noh, Adv. Mater. {\bf 30}, 1704777 (2018).

\bibitem{hotta} T. Hotta, Rep. Prog. Phys. {\bf 69}, 2061 (2006). 

\bibitem{mott} N. F. Mott, Rev. Mod. Phys. {\bf 40}, 677 (1968).

\bibitem{george} A. Georges, G. Kotliar, W. Krauth, and M. Rozenberg, Rev. Mod. Phys. {\bf 68}, 13 (1996).

\bibitem{kotliar} G. Kotliar, S. Y. Savrasov, K. Haule, V. S. Oudovenko, O. Parcollet, and C. A. Marianetti, Rev. Mod. Phys. {\bf 78}, 865 (2006).

\bibitem{slater} J. C. Slater, Phys. Rev. {\bf 82}, 538 (1951).

\bibitem{anderson} P. W. Anderson, Phys. Rev. {\bf 109}, 1492 (1958).
\bibitem{evers} F. Evers and A. D. Mirlin, Rev. Mod. Phys. {\bf 80}, 1355 (2008).

\bibitem{kim1} K. H. Kim, J. H. Jung, and T. W. Noh, Phys. Rev. Lett. {\bf 81}, 1517 (1998).
\bibitem{kim2} M. W. Kim, J. H. Jung, K. H. Kim, H. J. Lee, J. Yu, T. W. Noh, and Y. Moritomo, Phys. Rev. Lett. {\bf 89}, 016403 (2002).

\bibitem{brinkman} W. F. Brinkman and T. M. Rice, 
Phys. Rev. B {\bf 2}, 4302 (1970).

\bibitem{mo} S.-K. Mo, J. D. Denlinger, H.-D. Kim, J.-H. Park, J. W. Allen, A. Sekiyama, A. Yamasaki, K. Kadono, S. Suga, Y. Saitoh, T. Muro, P. Metcalf, G. Keller, K. Held, V. Eyert, V. I. Anisimov, and D. Vollhardt, Phys. Rev. Lett. {\bf 90}, 186403 (2003). 

\bibitem{panaccione} G. Panaccione, M. Altarelli, A. Fondacaro, A. Georges, S. Huotari, P. Lacovig, A. Lichtenstein, P. Metcalf, G. Monaco, F. Offi, L. Paolasini, A. Poteryaev, O. Tjernberg, and M. Sacchi, Phys. Rev. Lett. {\bf 97}, 116401 (2006).

\bibitem{eguchi}
R. Eguchi, M. Taguchi, M. Matsunami, K. Horiba, K. Yamamoto, Y. Ishida, A. Chainani, Y. Takata, M. Yabashi, D. Miwa, Y. Nishino, K. Tamasaku, T. Ishikawa, Y. Senba, H. Ohashi, Y. Muraoka, Z. Hiroi, and S. Shin, Phys. Rev. B {\bf 78}, 075115 (2008). 

\bibitem{rodolakis}
F. Rodolakis, B. Mansart, E. Papalazarou, S. Gorovikov, P. Vilmercati, L. Petaccia, A. Goldoni, J. P. Rueff, S. Lupi, P. Metcalf, and M. Marsi, Phys. Rev. Lett. {\bf 102}, 066805 (2009). 

\bibitem{sekiyama}
A. Sekiyama, H. Fujiwara, S. Imada, S. Suga, H. Eisaki, S. I. Uchida, K. Takegahara, H. Harima, Y. Saitoh, I. A. Nekrasov, G. Keller, D. E. Kondakov, A. V. Kozhevnikov, Th. Pruschke, K. Held, D. Vollhardt, and V. I. Anisimov, Phys. Rev. Lett. {\bf 93}, 156402 (2004).


\bibitem{akaki}
O. Akaki, A. Chainani, T. Yokoya, H. Fujisawa, T. Takahashi, and M. Onoda, Phys. Rev. B 56, 12050 (1997).


\bibitem{damascelli} 
A. Damascelli, Z. Hussain, and Z.-X. Shen, Rev. Mod. Phys. {\bf 75}, 473 (2003).

\bibitem{nakata} 
S. Nakata, M. Horio, K. Koshiishi, K. Hagiwara, C. Lin, M. Suzuki, S. Ideta, K. Tanaka, D. Song, Y. Yoshida, H. Eisaki, and A. Fujimori, npj Quantum Mater. {\bf 6}, 86 (2021).

\bibitem{fujimori} 
A. Fujimori, I. Hase, H. Namatame, Y. Fujishima, Y. Tokura, H. Eisaki, S. Uchida, K. Takegahara, and F. M. F. de Groot,
Phys. Rev. Lett. {\bf 69}, 1796 (1992).

\bibitem{uchida} 
S. Uchida, T. Ido, H. Takagi, T. Arima, Y. Tokura, and S. Tajima, 
Phys. Rev. B {\bf 43}, 7942 (1991).

\bibitem{marel}
D. van der Marel, H. J. A. Molegraaf, J. Zaanen, Z. Nussinov, F. Carbone, A. Damascelli, H. Eisaki, M. Greven, P. H. Kesm, and M. Li, 
Nature {\bf 425}, 271 (2003).

\bibitem{qazilbash}
M. M. Qazilbash, A. A. Schafgans, K. S. Burch, S. J. Yun, B. G. Chae, B. J. Kim, H. T. Kim, and D. N. Basov,
Phys. Rev. B {\bf 77}, 115121 (2008).

\bibitem{matsuura}
A. Y. Matsuura, H. Watanabe, C. Kim, S. Doniach, Z.-X. Shen, T. Thio, and J. W. Bennett, Phys. Rev. B {\bf 58}, 3690 (1998).

\bibitem{steve} S. M. Gilbertson, T. Durakiewicz, G. L. Dakovski, Y. Li, J.-X. Zhu, S. D. Conradson, S. A. Trugman, and G. Rodriguez, 
Phys. Rev. Lett. {\bf 112}, 087402 (2014).

\bibitem{dhaka} R. S. Dhaka, T. Das, N. C. Plumb, Z. Ristic, W. Kong, C. E. Matt, N. Xu, Kapildeb Dolui, E. Razzoli, M. Medarde, L. Patthey, M. Shi, M. Radovi\'{c}, and Jo\"{e}l Mesot, Phys. Rev. B {\bf 92}, 035127 (2015).


\bibitem{ronning}
F. Ronning, C. Kim, D. L. Feng, D. S. Marshall, A. G. Loeser,
L. L. Miller, J. N. Eckstein, I. Bozovicand, and Z.-X. Shen,
Science {\bf 282}, 5396 (1998).


\bibitem{kim3}
C. Kim, P. J. White, Z.-X. Shen, T. Tohyama, Y. Shibata, S.
Maekawa, B. O. Wells, Y. J. Kim, R. J. Birgeneau, and M. A.
Kastner, Phys. Rev. Lett. {\bf 80}, 4245 (1998).


\bibitem{vishik} I. M. Vishik, Rep. Prog. Phys. {\bf 81}, 062501 (2018).


\bibitem{huscroft}
C. Huscroft, M. Jarrell, Th. Maier, S. Moukouri, and A. N. Tahvildarzadeh, 
Phys. Rev. Lett. {\bf 86}, 139 (2001).

\bibitem{stanescu}
T. D. Stanescu and P. Phillips,
Phys. Rev. Lett. {\bf 91}, 049901 (2003).

\bibitem{senechal} 
David S\'{e}n\'{e}chal and A.-M. S. Tremblay, 
Phys. Rev. Lett. {\bf 92}, 126401 (2004).

\bibitem{wu} 
W. Wu, M. S. Scheurer, S. Chatterjee, S. Sachdev, A. Georges, and Michel Ferrero, Phys. Rev. X {\bf 8}, 021048 (2018).

\bibitem{he} Y. He, S.-D. Chen, Z.-X. Li, D. Zhao, D. Song, Y. Yoshida, H. Eisaki, T. Wu, X.-H. Chen, D.-H. Lu, C. Meingast, T. P. Devereaux, R. J. Birgeneau, M. Hashimoto, D.-H. Lee, and Z.-X. Shen, Phys. Rev. X {\bf 11}, 031068 (2021).



\bibitem{mayr} M. Mayr, G. Alvarez, C. S\'{e}n, and E. Dagotto, 
Phys. Rev. Lett. {\bf 94}, 217001 (2005)

\bibitem{anamitra} 
A. Mukherjee, N. D. Patel, S. Dong, S. Johnston, A. Moreo, and E. Dagotto, Phys. Rev. B {\bf 90}, 205133 (2014).


\bibitem{kumar} S. Kumar and P. Majumdar,  Eur. Phys. J. B {\bf 50}, 571 (2006).


\bibitem{kohn} W. Kohn, Phys. Rev. Lett. {\bf 76}, 3168 (1996).


\bibitem{salafranca} J. Salafranca, G. Alvarez, and E. Dagotto, 
Phys. Rev. B {\bf 80}, 155133 (2009).

\bibitem{jang} S. W. Jang, H. Sakakibara, H. Kino, T. Kotani, K. Kuroki and M. J. Han, Sci Rep {\bf 6}, 33397 (2016).


\bibitem{renner1} C. Renner, B. Revaz, J.-Y. Genoud, K. Kadowaki, and O. Fischer, Phys. Rev. Lett. {\bf 80}, 149 (1998).



\bibitem{ding} H. Ding, T. Yokoya, J. C. Campuzano, T. Takahashi, M. Randeria, M. R. Norman, T. Mochiku, K. Kadowaki, and J.
Giapintzakis, Nature {\bf 382}, 51 (1996).

\bibitem{loeser} A. G. Loeser, Z.-X. Shen, D. S. Dessau, D. S. Marshall, C. H. Park, P. Fournier, and A. Kapitulnik, Science {\bf 273}, 325 (1996).

\bibitem{norman} M. R. Norman, H. Ding, M. Randeria, J. C. Campuzano, T. Yokoya, T. Takeuchi, T. Takahashi, T. Mochiku, K. Kadowaki, and P. Guptasarma, Nature {\bf 392}, 157 (1998).


\bibitem{yoshida} T. Yoshida, X. J. Zhou, 
T. Sasagawa, W. L. Yang, P. V. Bogdanov, A. Lanzara, Z. Hussain,
T. Mizokawa, A. Fujimori, H. Eisaki, Z.-X. Shen, T. Kakeshita, and
S. Uchida, Phys. Rev. Lett. {\bf 91}, 027001 (2003).

\bibitem{kanigel1} A. Kanigel, M. R. Norman,
M. Randeria, U. Chatterjee, S. Souma, A. Kaminski, H. M. Fretwell, 
S. Rosenkranz, M. Shi, T. Sato, 
T. Takahashi, Z. Z. Li, H. Raffy, K. Kadowaki, D. Hinks, L. Ozyuzer, and
J. C. Campuzano, Nature Physics {\bf 2}, 447 (2006).

\bibitem{kanigel2} A. Kanigel, U. Chatterjee, M. Randeria, M. R. Norman, 
S. Souma, M. Shi, Z. Z. Li, H. Raffy, and J. C. Campuzano,
 Phys. Rev. Lett. {\bf 99}, 157001 (2007).

\bibitem{hasimoto} M. Hashimoto, I. M. Vishik, R.-H. He, T. P. Devereaux,
and Z.-X. Shen, Nature Physics {\bf 10}, 483 (2014).

\bibitem{dheeraj} D. K. Singh, S. Kadge, Y. Bang, and P. Majumdar
Phys. Rev. B {\bf 105}, 054501 (2022).

\bibitem{} P. Hansmann, N. Parragh, A. Toschi, G. Sangiovanni, and K. Held, New J. Phys. {\bf 16} 033009 (2014).

\end{thebibliography}
\end{document}